\documentclass[prl,superscriptaddress,twocolumn,letterpaper]{revtex4-1}
\usepackage{bm,graphicx,graphics,amsmath,amssymb,bm,epsfig,color}
\usepackage{euscript,tabularx}
\usepackage{longtable}
\usepackage{MnSymbol,wasysym,stmaryrd,ifsym,rotating,marvosym}
\usepackage{amsmath}
\usepackage{float}

\begin{document}

\title{The role of  damping rate amplitude in the synchronization of two coupled oscillators}

\author{A. Hamadeh}
\email{hamadeh@rhrk.uni-kl.de}
\affiliation{Fachbereich Physik, Technische Universitat Kaiserslautern, Kaiserslautern, Germany}

\author{A. Koujok} \affiliation{Fachbereich Physik, Technische Universitat Kaiserslautern, Kaiserslautern, Germany}
\author{I. Medlej} \affiliation{Southern University of Science and Technology, Shenzhen, China}
\author{P. Pirro} \affiliation{Fachbereich Physik, Technische Universitat Kaiserslautern, Kaiserslautern, Germany}
\author{S. Petit} \affiliation{Institut Jean Lamour, Universit\'e de Lorraine, UMR 7198 CNRS, 54506 Vandoeuvre-l\`es-Nancy, France}

\date{\today}

\begin{abstract}

We investigate the synchronization phenomenon between two Spin-transfer Torque Nano-Oscillators (STNOs) of different frequencies in two pillar systems under vortex configuration detunings or driving frequencies. The oscillators' mutual synchronization occurs through magnetic dipolar interaction. Our micromagnetic simulations show that an amplitude fluctuation referred to as $\Gamma_p$ has a significant impact on determining the synchronization frequency. The evolution of frequency and amplitude fluctuation rate in two different oscillator sizes versus external perpendicular field are compared and discussed. Our results reveal that the oscillator with lower $\Gamma_p$, referred to as the "Leader" oscillator, leads the synchronization process. As such, the "follower" oscillator adjusts its frequency as to that of the "Leader", thus achieving synchronization. We believe that taking $\Gamma_p$ into consideration can help in controlling synchronization frequencies in future building blocks of any network multi-array spintronics' devices.
  
\bigskip

\textbf{Keywords}: synchronization, dipolar coupling, nano-pillar oscillator, micromagnetic simulations.

\end{abstract}

\maketitle  

Synchronization is one valuable yet common phenomenon in non-linear oscillatory systems. Synchronization of biological, chemical and physical systems constitute vastly researched topics. For instance, these range from the synchronous behaviors observed in biological systems such as fireflies \cite{ermentrout1991adaptive}, to the synchronization of pendulum clocks \cite{rosenblum2003synchronization,kapitaniak2012synchronization,oliveira2015huygens}, all the way to the phase-locking of Spin-transfer Torque Nano-Oscillators (STNOs) studied by Kaka \cite{kaka2005mutual}, and Mancoff \cite{mancoff2005phase}. That stated, the synchronization of STNOs itself draws much attention in various fields such as bio-inspired oscillatory computing \cite{grollier2016spintronic,locatelli2014spin}, neuromorphic computing \cite{torrejon2017neuromorphic}, vowel recognition \cite{romera2018vowel} and neural networks \cite{cao2017fixed,leone2015synchronization,tang2011synchronization}. STNOs norm extensive scientific attention, as they constitute promising candidates for realization of physical oscillatory network arrays meeting all technical requirements such as room temperature operation, integration and scaling \cite{csaba2013computational,chen2016spin}.

Basically, we control the dynamics of the  magnetic layers in  STNOs via the spin transfer torque effect (STT) proposed by Slonczewski \cite{slonczewski1996current} and Berger \cite{berger1996emission}. STNOs are  composed of a thick magnetic layer (polarizing layer) with static direction of magnetization, a nonmagnetic spacer, then a thin magnetic layer (free layer). Magnetization oscillations in STNOs can be converted into microwave signals either through tunneling magnetoresistance (TMR), or giant magnetoresistance (GMR) \cite{reig2017giant}. Synchronization of STNOs can be achieved by various physical mechanisms such as electrical connection in series \cite{grollier2006synchronization,li2010global,lebrun2017mutual}, spin wave propagation \cite{kaka2005mutual,mancoff2005phase} and magneto-dipolar interaction \cite{belanovsky2012phase,araujo2015optimizing,hamadeh2014perfect,locatelli2015efficient,li2018selective}. 

It is well known that when two oscillators are synchronized, their oscillation frequencies are tuned. Hence, it is quite necessary to question the causes and key parameter(s) that determine their synchronization frequency. In other words, it is crucial to investigate the effect of coupling on the determination of this synchronization frequency. In this work, we study the synchronization of two dipolarly coupled circular pillar STNOs with magnetic vortex configurations in the upper (free) layers. Such synchronization process has been formerly shown to be efficient and inherent as emphasized in these works \cite{belanovsky2012phase,araujo2015optimizing,locatelli2015efficient}. Indeed, it is possible to excite different modes in STNOs by changing their geometrical and magnetic parameters, and tuning the bias magnetic field conditions. We will  consider only the case of vortices with opposite polarities and chiralities, which has already demonstrated experimentally the possibility of observing  synchronization \cite{locatelli2015efficient,li2018selective}. More specifically, the role of vortex polarity in the synchronization of nano-oscillators has been studied, reported and shown to be more efficient when polarities are opposite \cite{locatelli2015efficient}.

\begin{figure}[!h]
  \includegraphics[width=\columnwidth]{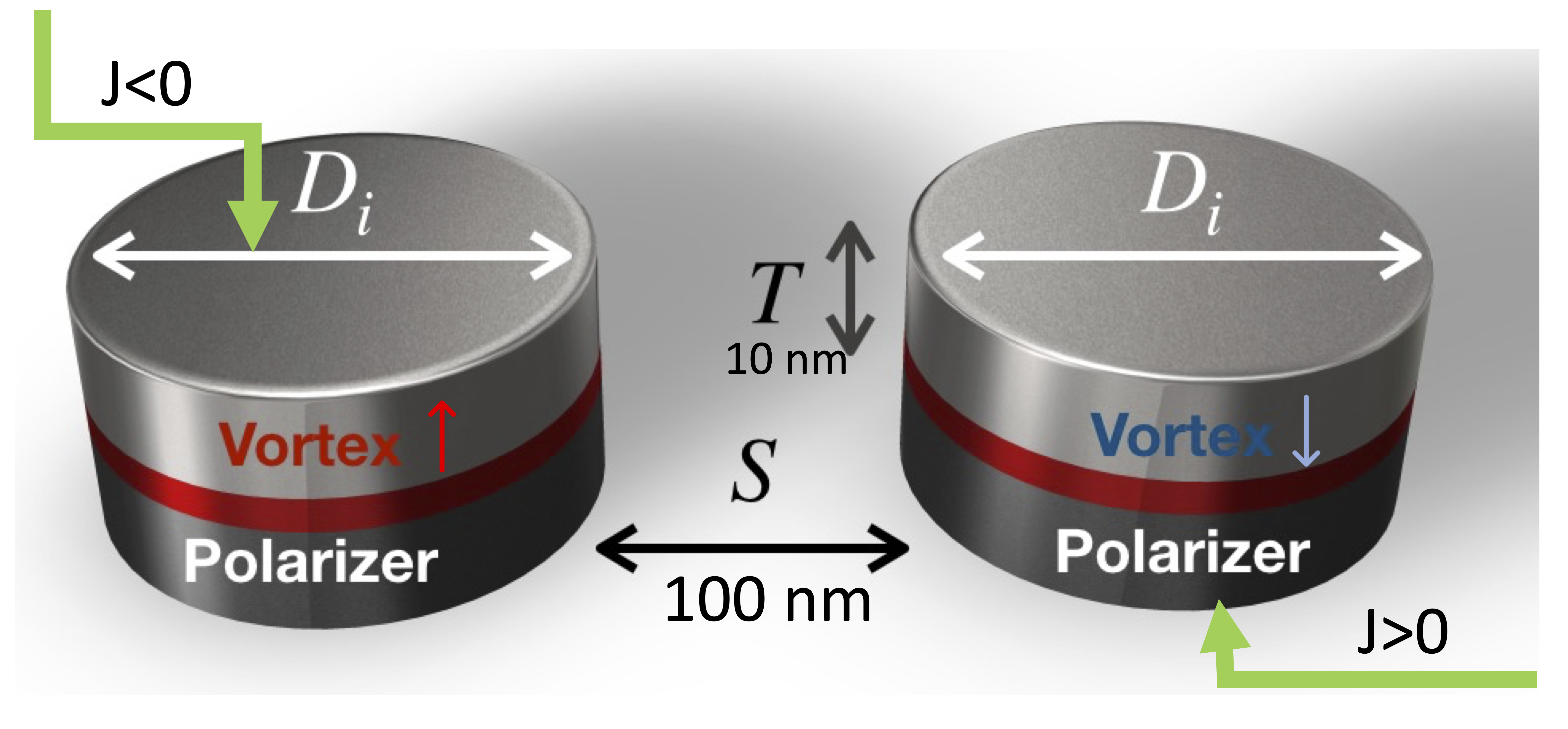}
 \caption{(Color online). Geometry of the studied system. The system is composed of two oscillators, each one is in a magnetic vortex configuration. The thickness and the separation are kept constant, while the discs' diameters are switched between Di=200 nm and Di=400 nm (with i=1,2). Polarities and chiralities are opposite in the two disks (See red/blue arrows for vortex polarities). Current density of same magnitude (J = 1.475 $\times$ 10$^{11}$ A/m$^2$) but opposite direction has been injected through each oscillator.}
  \label{fig:1}
\end{figure}

\begin{figure*}
\centering
 \includegraphics[width=12cm]{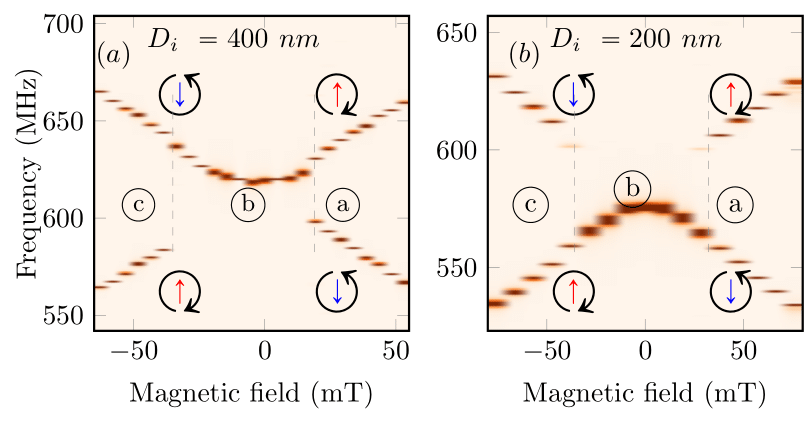}
  \caption{(Color online). Plots of the coupled STNOs' frequency dependence on the sweeping perpendicular applied field $H_\text{$\perp$}$ between three regions, namely {\textcircled{\small a}}
, {\textcircled{\small b}} and {\textcircled{\small c}}. The plots resemble discs of diameters (a) 400 nm and (b) 200 nm, respectively.}
  \label{fig:2}
\end{figure*}

Adopting the above mentioned efficient case, we show that changing the size parameters of the vortex-based STNOs, and tuning their frequencies with bias external perpendicular field will strongly modify their  synchronization behavior. We investigate numerically the synchronization properties for the selected combinations of vortex parameters, aiming to sort the best explanation as to achieving this synchronization.

In figure \ref{fig:1} we present our two pillar system. Each pillar consists of a free magnetic layer of (Py=Ni$_{80}$Fe$_{20}$) with thickness of 10 nm , a non-magnetic spacer layer, and a fixed polarizer (which produces a spin polarization). The circular multilayer stack was considered for two  different  sizes (200 nm and 400 nm). 

The computational studies are performed by means of micromagnetic simulations using the GPU (Graphics Processing Unit) based micromagnetic code Mumax \cite{ vansteenkiste2014design}, in addition to the software platform Aithericon \cite{aithericon}. The physical parameters of Py layers are: saturation magnetization Ms = 825 000 A/m, exchange constant
 A$_{ex}$ = 2.6 $ \times $ 10$^{-11}$ J/m, and damping $\alpha$ = 0.01. The vortices' parameters in each ferromagnetic nanopillar are referred to as P$_{1,2}$ (the vortex core (VC) polarity by vector, where $\uparrow$ stands for up and $\downarrow$ for down), and C$_{1,2}$ (the vortex Chirality by vector, where $\circlearrowleft$ stands for anticlockwise  and $\circlearrowright $ for clockwise). We consider two different sizes of  diameters with the same configurations for which self-sustained oscillations are achieved in both pillars. These configurations are of opposite polarities, with the cores' configurations corresponding to vortices moving in the opposite directions. In this case, the electrical connection is adapted according to the relative vortex core polarities in order to ensure self-sustained oscillations. Hereby, in order to ensure an opposite current sign in both pillars, we alimented these opposite polarity configurations using the parallel connection. The sweeping of the external perpendicular magnetic field  ($H${$_\perp$}) has an important role to tune the frequency. If the vortices have opposite polarities, then the variations of the frequency over $H${$_\perp$} given as dF/dH will have even opposite signs which makes it ideal. The sweeping allows us to  easily change the gap ($\Delta F$) between the two frequencies.The key idea is to take advantage of the slope dF/dH allowing us to work on this difference in frequency.

First, we investigate  the evolution of frequency peaks extracted 
from the  Fast Fourier Transform (FFT) of $M_{x}$  dynamics as a
function of $H${$_\perp$} (see Fig.\ref{fig:2}). The ground state of each disk  is a vortex  state. More specifically, the two vortices have been considered to have opposite   polarities ($P_1$ = $\uparrow$ and  $P_2$= $\downarrow$) and chiralities. This simulation is displayed using the color scale indicating  the amplitude integrated power generated by the system. Three different regions can be distinguished, namely \textcircled{a}, \textcircled{b} and \textcircled{c}. Starting with the case of discs with diameters $D_\text{i}$=400 nm (see Fig.\ref{fig:2} (a)), as the magnetic field decreases from 50mT to  20 mT (region \textcircled{a}), two signals are observed (see Fig. \ref{fig:2},a). The frequency  of each oscillator changes  with $H${$_\perp$}, that is the frequency of the vortex with polarity $\downarrow$  ($\uparrow$  ) increases (decreases) as $H${$_\perp$} decreases.  This region is called the subcritical  region where the coupling strength is not yet sufficient to synchronize the oscillators. As $\Delta F$  is reaching a threshold value of about 40 MHz at $H${$_\perp$}= 20 mT, the dynamical dipolar interaction leads to the synchronization of the vortex oscillations. In particular only one large amplitude peak subsists (see region \textcircled{b})), and reaches its maximum at $H${$_\perp$}=0. The synchronization frequency here is called a "higher branch" (HB) synchronization  frequency.   After synchronization,  as $H_{\perp}$ increases from -35 mT to -70 mT, the two signals reappear announcing  the ending of the synchronization. Now we take the case of two discs with  D$_i$=200 nm (see Fig.\ref{fig:2} (b)) , it  is clear that the synchronization occurs on the "lower branch" (LB). To understand the reason as to why the switching between HB/LB took place,  we need to find the physical factor that controls the STNOs' synchronization frequencies in both cases. Hence, it is necessary that magnetization oscillations should be considered and studied using the general model for nonlinear auto-oscillators \cite{slavin2009nonlinear}. The dynamics of a nonlinear oscillator can be described by the equation of the complex amplitude  (well described in ref. \cite{slavin2009nonlinear}). In the framework of the universal oscillator model \cite{grimaldi2014response}, it has been shown that the response  to small amplitude  perturbation of the vortex around its stable trajectory can be described by the damping rate for small power deviations from the stationary solution of the equation of the amplitude. This gives us the rate at which the system returns to the  stable trajectory after perturbation.

\begin{figure}
\centerline{\includegraphics[width=\linewidth]{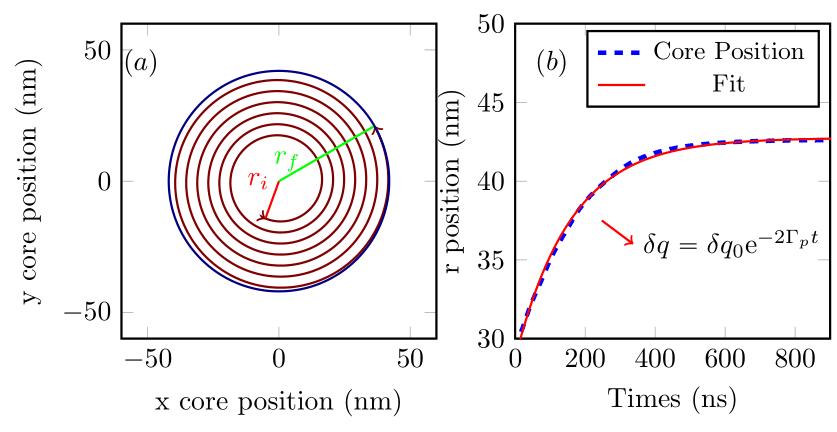}}
\caption{(Color online). (a) vortex core dynamics affected by amplitude perturbation $\delta q$ . (b) Core position as a function of time after perturbation, and the fit function that extracts the rate at which the system goes back to the stable trajectory.}
\label{fig:3}
\end{figure}

\begin{figure}[!h]
\centerline{\includegraphics[width=9cm]{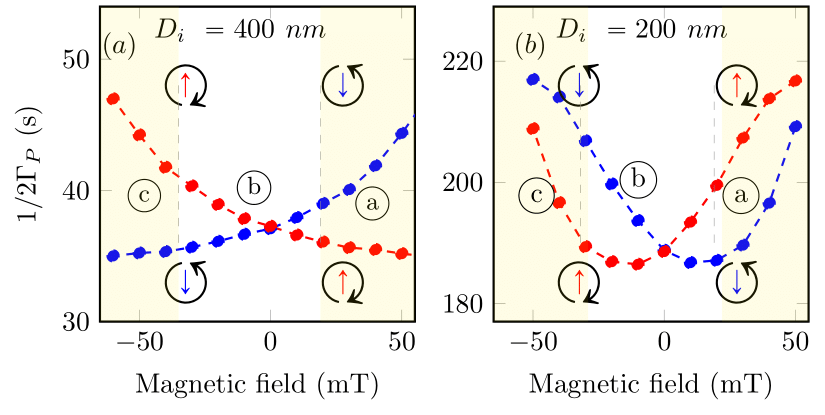}}
\caption{(Color online).Plots of 1/(2$\Gamma_p$) as a function of the perpendicular applied magnetic field in regions \textcircled{\small a}, \textcircled{\small b} and \textcircled{\small c} for discs of diameters (a) 400 nm and (b)200 nm, respectively.}
\label{fig:4}
\end{figure}

In Fig. \ref{fig:3} we present the  method of extracting the damping rate $\Gamma_P$ from simulation. In Fig. \ref{fig:3}(a) we have represented the VC's position in the x-y plane for a disc of diameter 200 nm. The figure indicates that the VC is shifted from the center  of the disc by a distance of ($r_f$-$r_i$ ). Fig. \ref{fig:3}(b) shows the results of the simulation as to how the VC's position changes with time. The vortex core continues to displace for a long distance from $r_i$  to reach $r_f$, at which it acquires a stable trajectory. Following the perturbation, the gyrotropic mode of the vortex relaxes back towards its equilibrium position with an amplitude relaxation rate $\Gamma_p$. The term 1/2$\Gamma_p$ is the  characteristic time of the relaxation of the amplitude and is given by :
\begin{equation}
\delta q(t) = \delta q_i e^{-2\Gamma_pt}
\end{equation} with $ \delta q$  being the  small amplitude deviations. The predicted results for a  perturbation affecting the magnetization dynamics of the vortex core are presented in  Fig. \ref{fig:4}. We first concentrate on  the quantitative analysis in Fig. \ref{fig:4}(a) for the   dependence of   1/2$\Gamma_p$ on $H${$_\perp$}  corresponding to $D_\text{i}$=400 nm. Here we  introduce  two types of oscillators, the first  one is called a "leader" oscillator, and the  other is the  "follower".  These nominations can be interpreted by the fact that the "follower" has  a stronger ability to adapt its frequency  to  the "leader". Focusing on the  region \textcircled{b}  which corresponds to the onset of  the "HB" synchronization, the oscillator with polarity $\uparrow$ follows the one with polarity $\downarrow$ for $H${$_\perp$} between 20 mT and 0 mT ($\Gamma_p$ ($\uparrow$)   $>$  $\Gamma_p$ ($\downarrow$)). But for   $H${$_\perp$}  between 0 mT and -20 mT  the  $\Gamma_p$'s  dependence on perpendicular field is reversed.  This can be interpreted by the  fact that the vortex with higher ($\Gamma_p$) has the ability to adapt its own frequency to the one with lower ($\Gamma_p$) . To gain further insight in our interpretation, we performed a similar study on discs with diameter size of 200nm. Here, the synchronization occurs on the lower branch "LB". From Fig. \ref{fig:4}(b), it is easy to conclude that the results oppose those of the first case. In other words, the roles between the "follower" and "leader" have been switched based on the behavior of $ \Gamma_p$'s dependence on perpendicular field.

Through micromagnetic simulations, we have unambiguously identified the role of the damping rate amplitude  in the synchronisation of two dipolarly coupled vortex-based STNOs. The oscillator with lower $\Gamma_\text{p}$ leads the synchronization process, while the second follows due to its ability to adjust its frequency to that of the first. This synchronization depends on the discs' geometries, polarities and applied magnetic field.  We believe that this finding might be very important for future designing strategies to select the synchronization frequency of arrays of STNOs.

\section*{Acknowledgements}
This research was funded by the European Research Council within the Starting Grant No. 101042439 "CoSpiN" and by the Deutsche Forschungsgemeinschaft (DFG, German Research Foundation) within the Transregional Collaborative Research Center—TRR 173–268565370 “Spin + X” (project B01).

\bibliography{references}
 
\end{document}